\documentclass[twocolumn,showpacs,aps,pre,
groupedaddress,amssymb,amsmath,nobalancelastpage]{revtex4}

\usepackage{graphicx}
\usepackage{longtable}

\begin{document}

\title{Early time evolution of Fre\`{e}dericks patterns generated from states of electroconvection}
\author{Denis Funfschilling*
 and Michael Dennin}
\address{Department of Physics and Astronomy, University of
California at Irvine, Irvine, California 92697-4575}
\address{*current address: Department of Physics and Astronomy,
University of California at Santa Barbara, Santa Barbara,
California}

\date{\today}

\begin{abstract}

We report on the early time ordering in a nematic liquid crystal
subjected to a sudden change in external ac electric field. We
compare time evolution for two different initial states of
electroconvection. Electroconvection is a highly driven state of a
nematic liquid crystal involving convective motion of the fluid
and periodic variations of the molecular alignment. By suddenly
changing either the voltage or the frequency of the applied ac
field, the system is brought to the same thermodynamic conditions.
The time ordering of the system is characterized by the evolution
of features of the power spectrum, including the average
wavenumber, total power, and shape of the the power spectrum. The
differences between the two classes of quenches are discussed, as
well as the possibility of scaling behavior during this initial
phase of domain growth.

\end{abstract}


\maketitle

Understanding systems driven far from equilibrium remains one of
the outstanding challenges of contemporary physics. At the heart
of the issue is the lack of any single principle that is
equivalent to the minimization of free energy that is applicable
in thermodynamic equilibrium \cite{CH93}. A subset of this larger
issue is the question of the transition between states of a system
after a sudden change in an external parameter, or a quench. The
general question of the behavior of systems after a quench is
often referred to as phase ordering or coarsening \cite{REV}, as
domains of the new steady state of the system order and grow.

Classically, phase ordering has been studied in the context of the
transition between two $\it equilibrium$ states. In this case,
minimization of a free energy plays a central role in
understanding the ordering process \cite{REV}. More recently,
transitions between $\it driven$ states of a system have gained
interest
\cite{EVG92,CM95,CB98,BV02,QM03,QM04,B04,PD01,KFD04,KID04}. In
this case, one considers the ordering of steady-states of a driven
system after the driving force has been changed. The free energy
does not play a role. However, there is growing evidence that the
dynamics of topological defects govern the phase ordering in both
equilibrium and nonequilibrium systems \cite{REV,HCSH02,KID04}.
One system that is useful for studying quenches in driven systems
is electroconvection \cite{PD01,KFD04,KID04}.

Electroconvection occurs when a nematic liquid crystal is placed
between two plates and an ac electric voltage is applied
perpendicular to the plates \cite{KP95,GP93}. A nematic liquid
crystal consists of long, rod-like molecules that on average are
aligned along a fixed axis, which is referred to as the director.
In the absence of electroconvection, the director is spatially
uniform. Above a critical voltage, the director develops a
periodic spatial variation, and there is an associated periodic
flow, or convection rolls. The most common geometry for
electroconvection is to have the director parallel to the glass
plates initially. An interesting case exists when the director is
initially perpendicular to the plates, also known as homeotropic
alignment \cite{HHHTK97,TBPK98}.

An important parameter that is used to characterize nematic liquid
crystals is the dielectric anisotropy \cite{GP93}. For an
anisotropic material, such as a nematic liquid crystal, the
dielectric constant is a tensor. The dielectric anisotropy is the
difference between the dielectric constant for the case where the
electric field and director are perpendicular and the case where
they are parallel. For materials with a negative dielectric
anisotropy in a homeotropic configuration, the initial transition
as a function of applied voltage is the Fre\`{e}dericks
transition, which is followed by the transition to
electroconvection \cite{HHHTK97,TBPK98}. The Fre\`{e}dericks
transition corresponds to the director developing a tilt relative
to its initial alignment perpendicular to the plates. This state
is an equilibrium state in which the angle of tilt is fixed for a
given value of the voltage. However, there is a degeneracy due to
the fact that the director tilt can adopt any azimuthal angle.
Regions with the same azimuthal angle are usually referred to as
Fre\`{e}dericks domains. Once the Fre\`{e}dericks transition has
occurred, there is a component of the director parallel to the
plates. Therefore, there is a second critical voltage at which
electroconvection will occur.

The Fre\`{e}dericks transition is independent of frequency;
whereas, the critical voltage for electroconvection is frequency
dependent. This allows for exploration of two different types of
quenches to the $\it same$ equilibrium state. With different
starting points, either a rapid change in frequency or voltage can
bring the system to the same final frequency and voltage, and
hence the same Fre\`{e}dericks state. However, the system
evolution to reach this final state is different. For a quench
down in voltage, the average director tilt angle has to relax to
the correct value. For a quench in frequency, the average tilt
angle of director should already have the appropriate value. In
both cases, the pattern present in the director field, charge
distribution, and flow field, due to electroconvection have to
relax.

In this paper, we report on a comparison of these two quenches. We
focus on the initial dynamics from the time of the quench until
the point where the spatial patterns are similar. The system's
evolution is characterized by a number of measures based on the
power spectrum of the images. It should be noted that it is also
interesting to consider the late-time evolution of the system. As
the Fre\`{e}dericks state is an equilibrium state, the late-time
evolution should be the same for both systems \cite{REV}. However,
large-scale imperfections in the system complicate studies of the
late-time dynamics, for which large aspects ratios are critical.
Therefore, the question of the late-time behavior will be the
subject of future work.

For these experiments, homeotropic cells of the nematic liquid
crystal N4 were used. The liquid crystal N4 is a eutectic mixture
of the two isomers of 4-methoxy-4'-n-butylazoxybenzene (${\rm
CH_3O-C_6H_4-NON-C_6H_4-C_4H_9}$ and ${\rm
CH_3OC_6H_4-NNO-C_6H_4-C_4H_9}$). It was obtained from EM
Industries (a Merck company), now EMD Chemicals Inc. \cite{EMI}.
The N4 was used without further purification. The method for
obtaining homeotropic alignment is described in detail in
Ref.~\cite{CFD05}. Briefly, a surfactant coating is made on ITO
coated glass using a Langmuir-Blodgett technique. A mixture of
43\% of N4 and 57\% arachidic acid (C20) diluted in chloroform is
spread at the air-water interface, forming a monolayer that was
compressed to 10~mN/m. We used C20 obtained from Sigma-Aldrich
with a quoted purity of $\geq 99\%$. It was used without further
purification. The pressure was held constant while ITO glass is
coated with 10 layers of the Langmuir monolayer. After the coating
of the surface, the glass is baked in an oven at a temperature of
$50\ {\rm ^{\circ}C}$. A $25\ {\rm \mu m}$ mylar spacer is placed
between the two ITO glasses. Two opposite sides of the cell are
sealed with epoxy. The cell is filled with N4 by capillary action.
The two remaining sides of the cell are sealed with 5 minute
epoxy.

The details of the experimental apparatus are described in
Ref.~\cite{D00}. Essentially, there is a temperature control stage
that holds the sample at a constant temperature within $\pm 5\
{\rm mK}$. The sample is illuminated from below and imaged from
above using standard shadowgraph techniques. A pair of
crossed-polarizers (one below the sample and one above the sample)
allow for simultaneous imaging of electroconvection patterns and
of Fre\'{e}dericksz domains. The use of ITO coated glass allows
for the application of an ac voltage to the sample.

\begin{figure} [htb]
\includegraphics[width=3.0in]{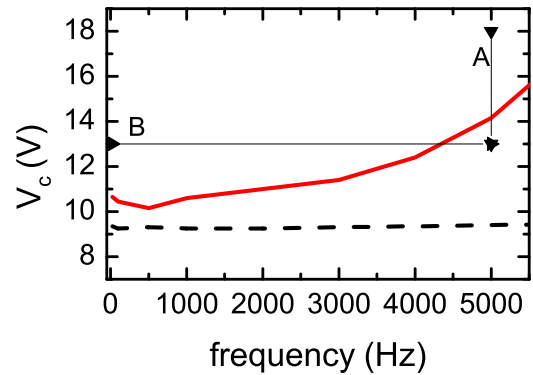}
\caption{A portion of the phase diagram for homeotropic N4 as
reported in Ref.~\cite{CFD05}. The dashed curve is the critical
voltage for the Fre\'{e}dericksz transition. The solid curve is
the critical voltage for the transition to electroconvection. The
horizontal and vertical lines connect the starting and ending
point of the two quenches. }
\end{figure}

The general state diagram for homeotropic samples of N4 is
reported in Ref.~\cite{CFD05}. There is the expected
Fre\`{e}dericks transition at approximately $V_{cF} = 9.3\ {\rm
V}$, independent of applied frequency. The transition to
electroconvection is frequency dependent. The solid black
horizontal and vertical lines with triangular endpoints in Fig.~1
illustrate the two types of quenches that are reported on in this
paper. The solid curve is the transition to electroconvection, and
the dashed black curve is the critical voltage for the
Fre\`{e}dericks transition (both curves were reported in
Ref.~\cite{CFD05}). Both quenches are selected to have the same
final equilibrium conditions of applied voltage and frequency
within the regime where Fre\`{e}dericks domains exist, but
electroconvection does not. The two starting points represent
different initial states for the system. The quench labelled A
starts from a chaotic electroconvection state and uses a change in
voltage. Physically, this means the tilt of the director must
change as part of the quench. The path labelled B starts from a
regular state of electroconvection and involves a change in
frequency. In this case, the distance from the Fre\`{e}dericks
critical voltage is not changed. Therefore, the details of the
initial evolution should be different.

For discussing these quenches, it is important to define a few
parameters. We will use $\epsilon = (V/V_c)^2 - 1$ to refer to the
distance from the electroconvection critical voltage ($V_c$) at a
fixed frequency. Therefore, since all of the quenches end at a
frequency $f = 5000\ {\rm Hz}$ and a rms voltage of $V = 13.0\
{\rm V}$, the final point for the quenches is $\epsilon = -0.156$
and $f = 5000\ {\rm Hz}$. The two starting points are $\epsilon =
0.618$ with $f = 5000\ {\rm Hz}$ and $\epsilon = 0.490$ with $f =
25\ {\rm Hz}$.

To provide a framework for the two quenches, we have characterized
the final state both by making quenches from below the
Fre\`{e}dericks critical voltage and by slowing stepping to the
final state. These studies involving increasing the voltage
established the existence of large-scale spatial inhomogeneities
in the Fre\`{e}dericks domains. Essentially the same spatial
pattern is obtained either by small steps in voltage or by
considering the late-time state after a large change in voltage.
The spatial inhomogeneities are most likely caused by small
defects in the aligning layer that result in the pinning of
defects in the director organization. Because of the large-scale
nature of the inhomogeneities, we do not expect them to play a
significant role in the early-time evolution of the system.
However, they do prevent detailed studies of the {\it late-time}
ordering at this point, and the influence of these domains can not
be completely ruled out for the early time evolution.

For each of the quenches, we measure a series of 128 images taken
1~s apart. We characterized the images using the spatial Fourier
transform of each image. We focus on the square of the modulus of
the Fourier transform of each image (the power spectrum,
$S(k,\theta)$, where $k$ is the wavenumber of interest). The power
spectra of 20 jumps are averaged image by image to improve the
statistics. There is a 15 minute wait between jumps. Using $S(k,
\theta)$, we measure the total power $P =
\int_0^{2\pi}\int_0^{\infty}S(k,\theta)kdkd\theta=2\pi\int_0^{\infty}S(k)kdk$,
where $S(k)$ is the azimuthal average of $S(k,\theta)$. From this
measure, we confirmed that the total power is essentially constant
throughout the time evolution, allowing comparison between
features of the power spectra at different times. The average
wavenumber,
\begin{equation}
<k>=\frac{2\pi\int_0^{\infty}S(k)k^2dk}{P}
\end{equation}
is used as a measure of the typical length scale in the system.
The rate of growth of this length scale can be compared for the
two quenches using the time-dependence of the average wavenumber.
Because we are looking at early time dynamics, it is not
necessarily expected that a scaling regime will exist. However, we
tested for scaling of the power spectrum of the form $S(k,t) = L^d
g(kL)$, where $L$ represents a typical length scale for the system
and $d$ is the spatial dimension of the pattern. For our case, $d
= 2$, and we take $L = <k>^{-1}$. Finally, in the study of phase
ordering, Porod's Law \cite{REV,JSS89} states that in the limit of
large wavelength ($kL >> 1$), $S(k,t) \sim L^{-n}k^{-(d+n)}$. In
this case, $d = 2$ and $n$ is the dimension of the vector order
parameters that describes the system. For a nematic liquid
crystal, one expects $n = 2$. In this sense, Porod's Law relates
the behavior of the long wavelength tail of the power spectrum to
the dominate defects in the system \cite{REV}.

\begin{figure} [htb]
\includegraphics[width=3.2in ]{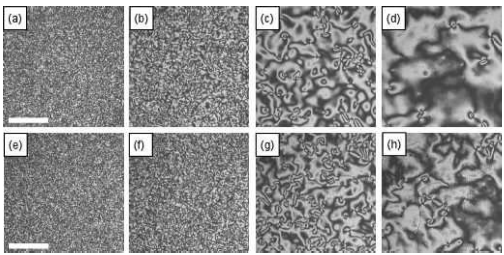}
\caption{Series of images for the frequency (a - d) and voltage (e
- h) quenches. The white bar in the images represents 1~mm. The
images are taken at 1~s, 4~s, 32~s, and 128~s after each quench.}
\end{figure}

Figure 2 presents the typical time evolution of the system after a
quench in voltage. Images 2(a) - (d) are for a frequency quench,
and images 2(e) - (h) are for a voltage quench. The images are
taken at 1~s, 4~s, 32~s, and 128~s after the quench and illustrate
the ordering that occurs. The images suggest that the voltage
quench evolves slower than the frequency quench. To quantify this
difference, we used the time dependence of $<k>$.

\begin{figure} [htb]
\includegraphics[width=3.0in]{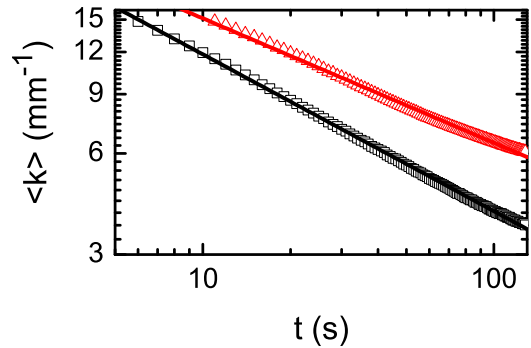}
\caption{Plot of the average wavenumber versus time for each of
the quenches. The frequency quench is represented by squares and
the voltage quench is represented by triangles. The solid line in
each case is a fit to a power law. The exponent for the frequency
quench is -0.47 and for the voltage quench, it is -0.37.}
\end{figure}

The main result of this paper is shown in Fig. 3. Plotted in Fig.
3 is the average wavenumber, $k$, versus time for both quenches.
The most dramatic feature of this result for the frequency quench
is the agreement with power law growth over the entire time range.
The voltage quench is also consistent with power law growth for
the time range plotted in Fig. 3. By comparing the exponents in
the power law (-0.47 for the frequency quench and -0.37 for the
voltage quench), one finds that the system orders faster after a
frequency quench than it does after a voltage quench.

\begin{figure} [htb]
\includegraphics[width=3.0in]{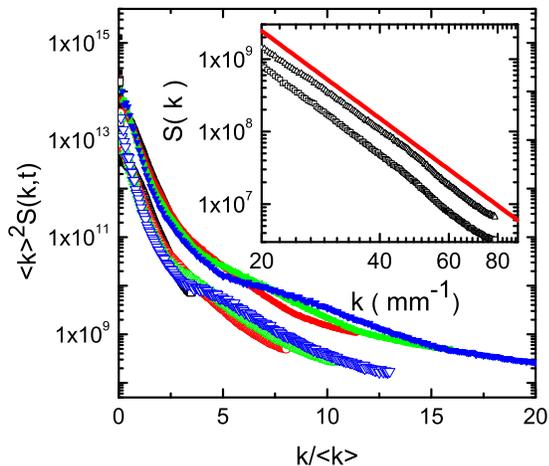}
\caption{Plot of the scaled power spectra versus the average
wavenumber for the frequency quench (solid symbols) and for the
voltage quench (open symbols). The power spectra for 2~s, 32~s,
64~s, and 128~s after the quench is shown for each case. The
insert shows the long time tail for 16~s after the frequency
quench (squares) and the voltage quench (triangles). The solid
line represents $k^{-4}$ and is provided as a guide to the eye.}
\end{figure}

The existence of power law behavior for the evolution of the
average wavenumber suggests that the system might be in a scaling
regime. The limited time of observation makes it difficult to
establish this with any certainty. However, as a test of scaling,
we considered the behavior of the full power spectra. The results
for scaling the azimuthally averaged power spectra are shown in
Fig. 4. The results are consistent with a scaling of the power
spectra, but are not conclusive. Perhaps more important than the
overall behavior of the spectra is the behavior at large
wavenumber. This is shown in the insert of Fig. 5 for a time 16~s
after the quench. The long wavelength part of the power spectra is
consistent with a power law that is independent of the quench
type: $S(k) \sim k^{-4}$ for large $k$. This result is in
agreement with the expected value for ordering of a nematic liquid
crystal in two dimensions \cite{REV}.

To summarize, the two quenches started from different values of
the average tilt of the director. Because the average wavenumber
was consistent with a power law, the different rates of phase
ordering could be quantified. The frequency quench (with the same
initial and final tilt) exhibited faster ordering than the voltage
quench. This is consistent with the different intrinsic time
scales for electroconvection. The two main relaxation times are
the charge relaxation time (on the order of $10^{-3}\ {\rm s}$ for
this system) and the director relaxation time (on the order of 1~s
for this system). Because the voltage quench should be dominated
by director relaxation at early times as the system equilibrates
to the equilibrium tilt angle, it is not too surprising that this
quench has slower dynamics.

Of interest for future work is the behavior of the exponents that
correspond to the power law growth of the average wavenumber. For
this system, the order parameter is not conserved, so one would
expect an exponent of $1/2$. This work was a focused study that
compared frequency quenches and voltage quenches. For the
frequency quench, the scaling of the average wavenumber was close
to $1/2$ (0.47), but for the voltage quench, it was significantly
smaller (0.37). It should be noted that for the voltage quench, by
considering different time windows (0.1 s, 0.5 s, and 1 s), we
observed apparent exponents for the evolution of $<k>$ of 0.28,
0.35, and 0.37, respectively. For the frequency quenches, the
exponents were essentially constant. This suggests two things. The
time dependence of the evolution of the two systems may become
similar in the long-time limit, if the trend for the voltage
quench continues. If that turns out to be true, both systems obey
the expected growth laws for an equilbrium system at late times.
However, more uniform systems and longer times are needed to
confirm this behavior, with a focus on the voltage quenches for
which the behavior is the most dramatic. Additionally, the impact
of the type of quench can be further explored using mixed quenches
where both the frequency and voltage is varied.

\begin{acknowledgments}

This work was supported by NSF grant DMR-9975479 and PRF
39070-AC9.

\end{acknowledgments}


\end{document}